\documentclass[preprint,showpacs,aps,preprintnumbers,amsmath,amssymb,
nofootinbib]{revtex4}
\usepackage{epsfig}
\begin{document}

\begin{flushright}
\end{flushright}


\newcommand{\be}{\begin{equation}}
\newcommand{\ee}{\end{equation}}
\newcommand{\bea}{\begin{eqnarray}}
\newcommand{\eea}{\end{eqnarray}}
\newcommand{\bers}{\begin{eqnarray*}}
\newcommand{\eers}{\end{eqnarray*}}
\newcommand{\nn}{\nonumber}
\newcommand\un{\cal{U}}
\def\Apa{A_\parallel}
\def\Ape{A_{\perp}}
\def\lp{\lambda^\prime}
\def\ll{\Lambda}
\def\mb{m_{\Lambda_b}}
\def\ml{m_\Lambda}
\def\s1{\hat s}
\def\ds{\displaystyle}
\def\s{\smallskip}
\def\l{\hspace*{0.05cm}}
\def\esp{\hspace*{1cm}}

\title{\large Unparticle effect on $B_s - \overline{B}_s$
mixing and its  implications for $B_s \to J/\psi \phi,~ \phi \phi $
decays }
\author{R. Mohanta$^1$  and A. K. Giri$^2$  }
\affiliation{$^1$ School of Physics, University of Hyderabad,
Hyderabad - 500 046, India\\
$^2$ Department of Physics, Punjabi University,
Patiala - 147 002, India}

\begin{abstract}
We study the effect of unparticle stuff on $B_s - \overline{B}_s$
mixing and consider possible implications of it for the decay
modes $B_s \to J/\psi \phi$ and $\phi \phi$. We find that due to the
new contributions from  the unparticles the $B_s - \overline{B}_s$
mixing phase could be observable at the LHC along with the possible
sizable CP asymmetry parameters $S_{\psi \phi( \phi \phi)}$ in $
B_s \to J/\psi \phi(\phi \phi)$ decay modes.
\end{abstract}

\maketitle

The standard model (SM) has been found to be very successful in
explaining the data up to the electroweak scale but still we believe
that it is the low energy manifestation of some beyond the standard
model scenario, which exists at high energy, the form of which is
not yet known. In the literature, there exist various
beyond the standard model scenarios which will
be tested in the upcoming experiments. Whereas, an interesting
and very much appealing idea has been proposed recently by Georgi
\cite{georgi} regarding the existence of some non-trivial scale
invariant hidden sector. Since the conventional particles are not described
by the scale invariant theory Georgi termed the physics described by
the scale invariant sector as the ``unparticle physics''.

In reality, the SM is not scale invariant and contains mostly
particles having nonzero mass but a scale invariant theory, if it
exists, can only have massless particles. It could be
possible that the SM fields at high energy might be scale invariant
but the scale invariance has to be broken at least at or above the
electroweak scale. Let us assume that the lack of scale invariance
of the SM is retained up to the  high energy scale  and further imagine that
there exist scale invariant fields at a higher scale above TeV with
a nontrivial infrared fixed point, termed as Banks-Zaks (${\cal{BZ}}$)
fields. Thus, the high energy theory contains both the SM fields and
the $B{\cal Z}$ fields. They interact via the exchange of particles
of large mass $M_{\un}$ which can generically be written as 
\bers
\frac{1}{M_{\un}^k} O_{SM} O_{\cal{BZ}}\;, 
\eers
where $O_{SM}$ is
the operator of mass dimension $d_{SM}$  and 
 $O_{\cal{BZ}}$ is the operator of mass dimension $d_{\cal{BZ}}$
 made out of SM  and ${\cal{BZ}}$ fields respectively.  At some scale
$\Lambda_{\un}$ the renormalizable
couplings of the ${\cal BZ}$ fields cause  dimensional
transmutation.  Below this scale ${\cal{BZ}}$ operators match
onto unparticle operators leading to a new set of interactions 
\bers
C_{\un}\frac{\Lambda_{\un}^{d_{\cal{BZ}}-d_{\un}}}{M_{\un}^k} O_{SM}
O_{\un}\;,
 \eers
where $C_{\un}$ is a coefficient function in the low energy effective theory
and $O_{\un}$ is the unparticle operator with scaling dimension
$d_{\un}$. Furthermore, $M_{\un}$ should be large enough such that
its coupling to the SM fields must be sufficiently weak, consistent
with the current experimental data. The production of these 
unparticles might be detectable by measuring 
the missing energy and momentum distribution in various processes
 \cite{georgi, kingman},
e.g., $t \to u+\un$, $e^-+e^+ \to \gamma +\un $, $Z \to q \bar q~ \un$,
etc.

Unparticle stuff  with scale dimension $d_{\un}$ looks like a
non-integral number $d_{\un}$ of invisible massless particles. Unparticle, if
exists, could couple to the standard model fields and consequently
affect the low energy dynamics. The effect of unparticle stuff on
low energy phenomenology has been explored in Refs.
\cite{kingman, ref2, lenz1}. One of the most interesting thing 
about unparticles is 
the existence of the peculiar  CP conserving phases in their
propagators
in the time like region, which lead to interesting CP violation phenomena.
For example, if nonzero direct CP asymmetry is found in the 
process $B^0 \to l^+ l^- $, it could be a direct signal 
of unparticle effects \cite{geng}. 

In this paper, we would like to see the effect of unparticle stuff
on the mass difference between the neutral $B_s$ meson mass
eigenstates ($\Delta M_s$) that characterizes the $B_s
-\overline{B}_s$ mixing phenomena. It is well known that flavor
changing $b \to s$ transitions are particularly interesting for new
physics searches. Among these $B_s - \overline{ B}_s$ mixing
plays a special role. In the SM, $B_s - \overline{ B}_s$ mixing
occurs at the one-loop level  by flavor-changing weak interaction
box diagrams and hence is very sensitive to new physics effects.
The effect of unparticle stuff in $B_s- \overline{B}_s$ 
mixing has also been recently investigated in Ref. \cite{lenz1}
where it is observed that large mixing phase could
be possible due to unparticle effects, 
in accordance with our findings.

In general $B_s - \overline{B}_s$ mass difference is defined as
$\Delta M_{B_s}  =  2 |M_{12}^s|= |\langle B_s^0| H_{eff}^{\Delta
B=2} | \overline{B}_s^0 \rangle|/M_{B_s}$, where  $H_{eff}^{\Delta B=2}$ is the
effective Hamiltonian responsible for the $\Delta B=2 $ transitions.
In the SM the mass difference  is given by \cite{mass}
\bea
 \Delta M_{B_s} = \frac{G_F^2 M_W^2}{6 \pi^2} M_{B_s} \hat \eta_B
\hat B_{B_s} f_{B_s}^2 |V_{tb} V_{ts}^*|^2 S_0(x_t)\;,
 \eea
where $\hat \eta_B$ is the QCD correction factor and $S_0(x_t)$ is
the Inami-Lim function \cite{lim} with $x_t=m_t^2/m_W^2$. In fact
the estimation of the SM value for $\Delta M_{B_s}$ contains large
hadronic uncertainties due to $ \hat B_{B_s} f_{B_s}^2$. Combining
the results of JL \cite{jl} and HPQCD \cite{hp} yields the $B_s$
mass difference as \cite {ball1}
 \be (\Delta M_{B_s})^{\rm SM}|_{\rm (HP+JL)QCD}=(23.4 \pm 3.8)~ {\rm
ps}^{-1}\;. \ee
Recently, Lenz and Nierste \cite{lenz2} updated the theoretical estimation
of the $B_s$ mass difference with value $(\Delta M_{B_s})^{\rm SM}=
(19.30 \pm 6.68)~ {\rm ps}^{-1}$ (for Set-I parameters) and 
$(\Delta M_{B_s})^{\rm SM}=
(20.31 \pm 3.25)~ {\rm ps}^{-1}$ (Set-II).

Experimentally, the D\O~ \cite{d0} and CDF \cite{cdf}  collaborations have
reported new results for the $B_s - \bar{B}_s $ mass difference \bea
&& 17 ~{\rm ps^{-1}} < \Delta M_{B_s} < 21~ {\rm ps^{-1}}~~~~~~~~90
\%~ {\rm C.L.}~
({\rm D \O })\nn\\
&& \Delta M_{B_s}=(17.77 \pm 0.10 \pm 0.07)~ {\rm
ps^{-1}}~~~~~~~~~({\rm CDF})\;. \eea Although the experimental
results appear to be consistent with the standard model prediction,
but they do not completely exclude the possible new physics effects
in $\Delta B=2$ transitions. In the literature, there have already
been many discussions both in model independent \cite{ball1, lenz2, np} and
model dependent way \cite{susy} regarding the implications of these
new measurements. In this work we would like to see the effect of
unparticle stuff on the mass difference of $B_s$ system and its
possible implications for the mixing induced CP asymmetries in $B_s
\to J/\psi \phi$ and $\phi \phi$ decay modes. In our analysis we use
the central value of (JL+HP)QCD results as the SM contribution and
the central value of CDF result as the experimental value for
$\Delta M_{B_s}$.

New physics contribution to the mixing amplitude $M_{12}^s$ can be
parameterized in the most general way as  \be
M_{12}^s=M_{12}^{\rm SM}+M_{12}^{\rm NP}=M_{12}^{\rm SM}(1-R e^{i
\phi})\;,\label{ms}
 \ee
where $ M_{12}^{\rm SM}$ and $ M_{12}^{\rm NP}$ are the SM and new
physics (NP) contributions, $R=|M_{12}^{\rm NP}/ M_{12}^{\rm SM}|$
and $\phi$ is the relative phase between them. It should be noted
here that since the SM contribution to $\Delta M_{B_s}$ is above the
present experimental value, we have explicitly made the NP
contribution to be negative in the last term of Eq. (\ref{ms}) so
that it will interfere destructively with the corresponding SM value
for $\phi=0$. Alternatively, one can also parameterize these
contributions as \be
 \sqrt{\frac{M_{12}^s}{M_{12}^{\rm SM}}} =r_s e^{i \theta_s}\;,
\ee which gives \be M_{12}^s=r_s^2 e^{2i \theta_s}~ M_{12}^{\rm
SM}\;. \ee Values of $r_s^2 \neq 1 $ and $2 \theta_s \neq 0$ would
signal new physics. These two sets of parametrization can be related
to each other by
\begin{eqnarray*}
r_s^2=\sqrt{1+R^2-2R \cos \phi}\;,~~~~{\rm and}~~~~ \tan 2 \theta_s
=\frac{-R \sin \phi}{1-R \cos \phi}\;.
\end{eqnarray*}
Now we proceed to see how unparticle stuff will affect the mixing
amplitude $M_{12}^s$. It should be noted that, depending on the
nature of the original ${\cal BZ}$ operator $O_{\cal BZ}$ and the
transmutation, the resulting unparticle may have different Lorentz
structure. In our analysis, we consider only two kinds of
unparticles i.e., scalar type  and vector type. 
Under the scenario that
the unparticle stuff transforms as a singlet under the SM gauge
group \cite{georgi},  
 the unparticles can couple to
different flavors of quarks and induce flavor changing neutral
current (FCNC) transitions even at the tree level.
Thus, the coupling of
these unparticles to quarks is given as \be
\frac{c_S^{q'q}}{\Lambda_{\un}^{d_{\un}}}\bar q'
\gamma_\mu(1-\gamma_5) q~
\partial^\mu O_{\un}+\frac{c_V^{q'q}}{\Lambda_{\un}^{d_{\un}-1}}\bar q'
\gamma_\mu(1-\gamma_5) q~ O_{\un}^\mu+h.c.\;, \label{cv}
\ee where
$O_{\un}$ and $O_{\un}^\mu$ denote the scalar and vector unparticle
fields and
$c_{S,V}^{q'q}$ are the dimensionless coefficients which in general depend on
different flavors. If both $q$ and $q'$ belong to up (down) quark
sector, FCNC transitions can be
induced by  the above effective interactions. Thus, the
unparticles mediate the $b \to s$  transitions in
the  $B_s -
\overline{B}_s$ mixing where they appear only as  propagators with
momentum $P$ and scale dimension $d_{\un}$.

The propagator for the scalar unparticle field is given as
\cite{georgi, kingman}
 \be \int d^4 x e^{i P \cdot x}\langle 0 | TO_{\un}(x)
O_{\un}(0)|0 \rangle = i \frac{A_{d_{\un}}}{2 \sin d_{\un} \pi}
\frac{1} {(P^2+i \epsilon)^{2-d_{\un}}}e^{-i\phi_{\un}}\;, \ee where
\be A_{d_{\un}}= \frac{16 \pi^{5/2}}{(2 \pi)^{2 d_{\un}}}
\frac{\Gamma(d_{\un}+1/2)}{\Gamma(d_{\un}-1)\Gamma(2d_{\un})}\;,
 ~~~~{\rm and} ~~~\phi_{\un}=(d_{\un}-2)\pi \;.\ee
Similarly the propagator for the vector unparticle is given by \be
\int d^4 x e^{i P \cdot x}\langle 0 | TO_{\un}^\mu(x) O_{\un}^\nu
(0)|0 \rangle = i \frac{A_{d_{\un}}}{2 \sin d_{\un} \pi}
\frac{-g^{\mu \nu} +P^\mu P^\nu/P^2} {(P^2+i
\epsilon)^{2-d_{\un}}}e^{-i \phi_{\un}}\;. \ee

From Eq. (\ref{cv}), one can easily see that the new effective
operators contributing to $B_s - \overline{B}_s$ due to vector/scalar type
unparticle exchange are given by \bea Q_{V-A} &=& \bar s
\gamma^\mu(1-\gamma_5)b ~\bar s \gamma_\mu(1-\gamma_5)b\;,
\nn\\
 Q_{S+P} &=& \bar s (1+\gamma_5)b~ \bar s (1+\gamma_5)b\;.
\eea
 Using the
vacuum insertion method, the matrix elements of these operators are
 given as \bea \langle \bar B_s | \bar s \gamma^\mu(1-\gamma_5)b
~\bar s \gamma_\mu(1-\gamma_5)b| B_s \rangle &= & \frac{8}{3}
f_{B_s}^2\hat B_{B_s} m_{B_s}^2  \;,
\nn\\
\langle \bar B_s| \bar s (1+\gamma_5)b~ \bar s (1+\gamma_5)b |B_s
\rangle &=& -\frac{5}{3} f_{B_s}^2\tilde B_{B_s} m_{B_s}^2\;. \eea
Thus, we get the new contributions to $M_{12}^s$ due to the
scalar/vector like unparticles  as \bea |M_{12}^{\un}|_{\rm
scalar}&=& \frac{5}{6}\frac{f_{B_s}^2 \tilde B_{B_s}}{ m_{B_s}}
\frac{A_{d_{\un}}}{2 |\sin d_{\un}\pi|}\left
(\frac{m_{B_s}}{\Lambda_{\un}} \right )^{2 d_{\un}} |c_S^{sb}|^2\;,
 \nn\\
 |M_{12}^{\un}|_{\rm vector}&=& \frac{1}{2}\frac{f_{B_s}^2 \hat
B_{B_s}}{ m_{B_s}} \frac{A_{d_{\un}}}{2| \sin d_{\un}\pi|}\left
(\frac{m_{B_s}}{\Lambda_{\un}} \right )^{2 d_{\un}-2}
|c_V^{sb}|^2\;.\label{lk}
 \eea
From the above equations one can see that the unparticle
contributions depend on three unknown parameters, namely, the
dimension of the unparticle fields $d_{\un}$, the scale
$\Lambda_{\un}$ and the couplings $c_{S,V}^{sb}$. Therefore, it is
not possible to constrain the new physics contributions unless we
fix some of these parameters. Now to obtain the constraint on the
coupling constants,  we fix the energy scale $\Lambda_{\un}$=1 TeV and
the scale dimension $d_{\un}$=3/2. We use the value of the decay constant
$f_{B_s} \sqrt{\hat B_{B_s}}=0.262$ GeV from Blanke {\it et al} 
in Ref. \cite{np} alongwith the relationship
between the  bag parameters \cite{lenz2} as $\tilde B_{B_s}=
\Big(m_{B_s}^2/(\overline{m}_b+\overline{m}_s)^2\Big)\hat B_{B_s} 
\approx 1.55~ \hat B_{B_s}$.
 Assuming that only scalar/vector type unparticles
contribute at a given time and the total contributions is given by
the unparticles one can obtain the upper bound on $c_{S,V}$ as \be
|c_S^{sb}| \leq 0.12 \;,~~~~{\rm and}~~~~~|c_V^{sb}| \leq 0.001\;.
\ee

Now to obtain the lower bound on $c_{S,V}^{sb}$, we assume that the
minimum value of the unparticle contribution is such that it will
just be sufficient to lower the SM contribution to the present
experimental value. Thus, from Eqs. (\ref{ms}) and (\ref{lk}) we
obtain the lower bounds as \be |c_S^{sb}| \geq 6.75 \times 10^{-2}
\;,~~~~{\rm and}~~~~~|c_V^{sb}| \geq 5.8 \times 10^{-4}\;. \ee

\begin{figure}[htb]
   \centerline{\epsfysize 2.5 truein \epsfbox{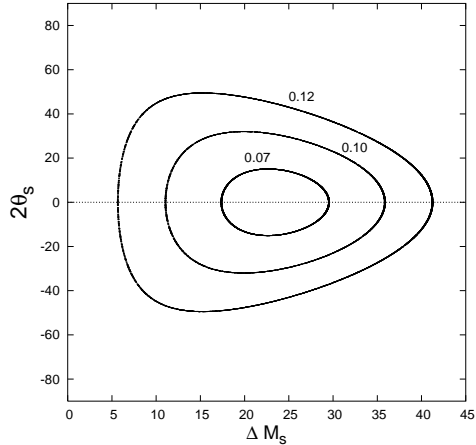}}
 \caption{
 Correlation plot between $\Delta M_{B_s}$ in ${\rm ps}^{-1}$ and
 $2 \theta_s$ in degree  for a representative set of
values of $|c_{S}^{sb}|$ as labelled in the plots.}
 \end{figure}
\begin{figure}[htb]
   \centerline{\epsfysize 2.5 truein \epsfbox{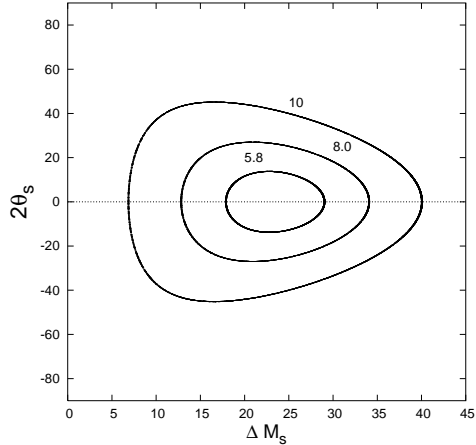}}
 \caption{ Same as Figure-1, with vector like unparticle contributions
where the constants $|c_{V}^{sb}|$ are in units of $10^{-4}$.}
 \end{figure}

Now in Figures-1 and 2, we plot $\Delta M_{B_s}$ versus
$ 2 \theta_s$ using Eq. (\ref{ms}) for some representative set of values
of $|c_{S,V}^{sb}|$ from the above range, where we have varied the weak phase
$\phi$ between 0 and $2 \pi$. From the figures it can be seen that
large value of mixing phase could be possible due to unparticle
effects.  Measurement of this phase in the upcoming experiments such as
LHC could imply an indirect evidence for the existence of unparticles.

Since large mixing phase is indeed
possible due to unparticle effect, now we  would like to look into
its possible implications in the mixing induced CP violation
in $B_s \to J/\psi \phi$. The $B_s \to J/\psi  \phi $ decay
channel is accessible at hadron colliders
where plenty of $B_s$ is expected to be produced. 
It is therefore considered as
one of the benchmark channels to be  studied at the LHCb
experiment. This decay  mode proceeds through the quark level transition
$b \to c \bar c s $ which is analogous to $B_d \to J/\psi K_S$.
However, the final state in  $B_s \to J/\psi \phi $ is not a CP
eigenstate but a superposition of CP odd and even states, which
 can be disentangled through an angular analysis of their decay
products \cite{dighe}.
Therefore, the mixing induced CP asymmetry in this mode  is expected
to give
\be
S_{J/\psi \phi}= -\sin 2 \beta_s\;,
\ee
where $ \beta_s \equiv {\rm
arg}(V_{tb} V_{ts}^*) \approx -1^\circ$.
Since this decay mode receives dominant contribution
from the color suppressed tree level transition $b \to c \bar c s$,
it is unlikely that new physics contribution
to the decay amplitude will significantly modify the SM
amplitude. Therefore, we will assume that the new physics contribution to
this decay amplitude is negligible and hence the
CP asymmetry
will be modified because of the new contributions to the mixing.
Thus, in the presence of NP the mixing-induced CP asymmetry can be
obtained as follows.
Assuming that there is no direct CP violation in this mode
one obtains
\bea
S_{J/\psi \phi} \sin \Delta M_s t & = & \frac{\Gamma (\overline{B}_s(t) 
\to J/\psi \phi )
-\Gamma(B_s(t) \to J/\psi \phi)}{\Gamma (\overline{B}_s(t) \to J/\psi \phi )
+\Gamma(B_s(t) \to J/\psi \phi)} \nn\\
&=&\frac{D~ {\rm Im}\left (\displaystyle{\frac{q}{p}} 
\rho_{\rm odd} \right )+{\rm Im} \left (
\displaystyle{\frac{q}{p}}  \rho_{\rm even} \right )}
{D F_{\rm odd}(t)+F_{\rm even}(t)} \sin \Delta M_s t
\eea
where
\bea
F_{\rm odd, even}(t)= \cosh \left (\frac{\Delta \Gamma_s t}{2}
\right )+
{\rm Re} \left [\frac{q}{p}\rho_{\rm odd, even} \right ]
\sinh \left ( \frac{\Delta\Gamma_s t}{2 }\right )\;,
\eea
with
\be
\rho_{\rm odd,even}=\frac{A(\overline{ B}_s \to J/\psi 
\phi)_{\rm odd, even}}
{A(B_s \to J/\psi \phi)_{\rm odd, even}}\;,~~~~~{\rm and}
~~~~~~D= \frac{|A_\perp|^2}{|A_\parallel|^2+|A_0|^2}\;.
 \ee
$\Delta \Gamma_s$ is the lifetime difference between heavy and light
$B_s$ eigen states. 
Thus, we get
\be
S_{J/\psi \phi}= \frac{(1-D) \sin 2 |\beta_s|}{(1+D)\cosh(\Delta \Gamma_s
t/2)+(1-D)\cos 2 \beta_s \sinh(\Delta \Gamma_s t/2)}\;.
\ee
Taking the limit $\Delta \Gamma_s \to 0$ and
scaling out the CP odd fraction we obtain
\be
S_{J/\psi \phi}^\prime= \frac{S_{J/\psi \phi}}{1-2 f_\perp}
= \sin (2 |\beta_s|-2 \theta_s)\;,
\ee
where $f_\perp= |A_\perp|^2/(|A_0|^2+|A_{\parallel}|^2+|A_\perp|^2)$.
Now  plotting $S_{J/\psi \phi}^\prime$ versus the new 
mixing phase $  \theta_s$ in figure-3,
we see that large CP violation could be possible in this mode.

\begin{figure}[htb]
   \centerline{\epsfysize 2.0 truein \epsfbox{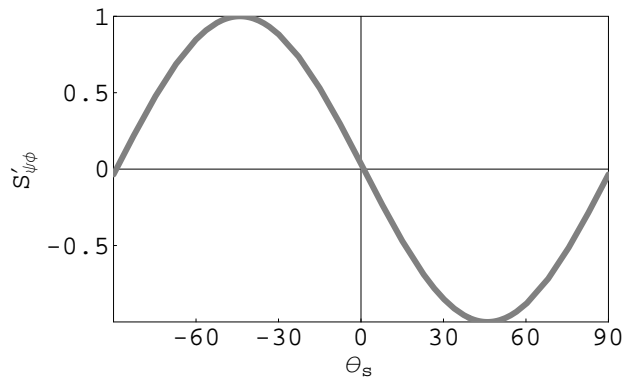}}
 \caption{The variation of $S_{J/\psi \phi}^\prime$ versus
the mixing
$ \theta_s$ in degree.}
  \end{figure}

Thereafter, we consider another decay channel $B_s \to \phi \phi$ which is a
pure penguin induced  process  and proceeds through the quark level
transition $b \to s \bar s s $. Assuming the top quark
dominance in the loop, the mixing induced
CP asymmetry in the SM turns out to be identically zero
because the weak phase in the mixing and in the ratio of decay amplitudes
exactly cancel each other. Since the dominant SM contribution arises
at the one-loop level it is expected that this decay channel may
receive  new contribution from NP in its decay amplitude, 
unlike the $B_s \to J/\psi
\phi$ process. Therefore, we are interested to see how $S_{\phi \phi}$ will
be modified due to the unpraticle contributions in its decay
amplitude.

To see the effect of NP in the decay amplitude we first consider the
SM amplitude. In general the decay mode $B_s \to \phi \phi$ can be described
in the helicity basis, where the amplitude for the helicity
matrix element can be parametrized as~\cite{Kra92}
\begin{eqnarray}
H_{\lambda}&=& \langle \phi (\lambda) \phi(\lambda)|{\cal H}_{eff} |B_s
\rangle\nonumber\\
&=& \varepsilon_{1 \mu}^* (\lambda) \l \varepsilon_{2
\nu}^* (\lambda) \left [ a g^{\mu \nu} + \frac{b}{m_\phi^2} p^{\mu}
p^{\nu} + \frac{i c}{m_\phi^2} \epsilon^{\mu \nu \alpha \beta} p_{1
\alpha} p_{\beta} \right ]\;,
\label{hlam}
\end{eqnarray}
where $p$ is the $B_s$ meson momentum and $\lambda =0, \pm 1$ are
the helicity of  the
$\phi$ mesons. In the above expression $p_i$ and $\varepsilon_i$
($i=1,2$) stand for their
momenta and polarization vectors of the two $\phi$ mesons respectively.
Furthermore, the three invariant amplitudes $a$, $b$, and $c$ are related to
the helicity amplitudes by
\begin{equation}
H_{\pm 1} = a \pm c \l \sqrt{x^2 - 1}\;,
\esp
H_0 = - a x - b \l (x^2 - 1)\;,
\label{a}
\end{equation}
where $x =(p_1 \cdot p_2)/m_\phi^2 = (m_B^2 -2 m_\phi^2)/2 m_\phi^2$.

The corresponding decay rate using the helicty basis amplitudes can be given as
\begin{equation}
\Gamma = \frac{p_{cm}}{8 \pi m_{B_s}^2} \biggr( |H_0|^2+|H_{+1}|^2 +|H_{-1}|^2
\biggr)\;,
\end{equation}
where $p_{cm}$ is the magnitude of c.o.m. momentum of the outgoing $\phi$
mesons.

The  amplitudes in  transversity and helicity basis are related to each other
through the following relations
\begin{eqnarray}
A_{\bot} \l = \l \frac{H_{+1} - H_{-1}}{\sqrt{2}}, \esp
A_{\|} \l = \l \frac{H_{+1} + H_{-1}}{\sqrt{2}}, \esp A_0 \l =
\l H_0 \label{cb}.
\end{eqnarray}

The SM amplitude for the process $\overline{B}_s \to \phi \phi $ can be
represented in the  factorization approach as
\bea
A(\overline{ B}_s \to \phi \phi)=-\frac{G_F}{\sqrt{2}}V_{tb}V_{ts}^* 2~\left[
a_3+a_4+a_5-\frac{1}{2}(a_7+a_9+a_{10}) \right] X\;,
 \label{A}
 \eea
where
\be
X \equiv \langle \phi(\varepsilon_2, p_2)|\bar s \gamma_\mu (1-\gamma_5)
s | 0 \rangle\langle \phi(\varepsilon_1, p_1)|\bar s \gamma^\mu
(1-\gamma_5)
b | \overline{ B}_s(p) \rangle \label{fc}\ee
is the factorizable hadronic matrix element and
$a_i$ are the QCD coefficients.
In the factorization approximation, the factorized matrix element
$X$ (Eq. (\ref{fc})) can be written, in general, in terms of form factors
and decay constants which are defined as
\begin{eqnarray}
\langle \phi (\varepsilon_2,p_2) | V_\mu  | 0 \rangle & = 
& f_\phi \, m_\phi \,
\varepsilon^\ast_{2 \mu}, \nonumber \\
\langle \phi (\varepsilon_1, p_1) | V_\mu  |  B_s (p) \rangle & =
& \frac{2}{m_{\phi} +
m_{B_s}} \; \epsilon_{\mu\nu\alpha\beta} \, \varepsilon_1^{\ast\,\nu} \,
p^\alpha {p_1}^\beta \, V(q^2), \nonumber \\
\langle \phi (\varepsilon_1, p_1) | A_\mu 
|  B_s (p) \rangle & = &
-i\,
\frac{2\, m_\phi (\varepsilon_1^\ast\cdot q)}{q^2}\;
q_\mu\; A_0(q^2)
-i\, (m_{\phi} + m_{B_s}) \left[\varepsilon_{1 \mu}^\ast
- \frac{(\varepsilon_1^\ast
\cdot q)}{q^2}\; q_\mu \right] A_1 (q^2) \nonumber \\
&  +& i \left[(p + p_1)_\mu\, - \frac{(m_{B_s}^2-m_{\phi}^2)}{q^2}
q_\mu \right] \frac{(\varepsilon_1^\ast \cdot q)}{m_{\phi} + m_{B_s}} 
A_2(q^2)\;,
\end{eqnarray}
where $V_\mu$ and $A_\mu$ are the corresponding vector and axial
vector quark currents and  $q=p-p_1$ as the momentum transfer.
In this way the invariant amplitudes $a$, $b$, and $c$ read as
\begin{eqnarray}
a & = & i ~C_{eff}\,f_\phi~ m_\phi \, ( m_{B_s} + m_{\phi} )\,
A_1^{B_s\to \phi}(m_\phi^2),
\nonumber \\
b & = & -i~ C_{eff}\,f_\phi~ m_\phi \, \left(\frac{2\, m_{\phi}^2}
{m_{B_s} + m_{\phi}} \right)  A_2^{B_s\to \phi}(m_\phi^2), \nonumber \\
c & = & -i~ C_{eff}\, f_\phi ~m_\phi \, \left(\frac{2\, m_{\phi}^2}
{m_{B_s} + m_{\phi}} \right)  \, V^{B_s\to \phi}(m_\phi^2),
\label{peng}
\end{eqnarray}
where
\begin{equation}
C_{eff} = -\frac{G_F}{\sqrt{2}} \; V^\ast_{ts}\, V_{tb}2 \left[a_3 + a_4
+ a_5 - \frac{1}{2} ( a_7 + a_9 + a_{10} )\right]\,.
\label{ceffp}
\end{equation}

The values of the QCD improved effective
coefficients $a_i$ can be found in Ref. \cite{cheng}. Now
substituting the values of $a_i$ for $N_C$=3, from Ref. \cite{cheng},
the value of the form factor $V^{B_s \to \phi}(m_\phi^2)=$ 0.461,
$A_1^{B_s \to \phi}(m_\phi^2)=0.317$, $A_2^{B \to \phi}(m_\phi^2)=$ 0.245
are obtained using the LCSR approach \cite{ball},
and using
the $\phi$ meson decay constant $f_{\phi}=$ 0.231 GeV,
$|V_{tb}V_{ts}^*|=41.3 \times 10^{-3}$ and $\tau_{B_s}=
1.466 \times 10^{-12}$ sec \cite{pdg}, we obtain
the branching ratio in the SM as
\be
BR^{SM}(\bar B_s \to \phi \phi)=10.4
\times 10^{-6}.
\ee
which appears to be consistent with the experimental
value $BR(B_s \to \phi \phi)=(14_{-7}^{+8}) \times 10^{-6}$ \cite{hfag}. 
But still one cannot
rule out the possibility of NP in the decay amplitude as we need
the measurement of CP violating parameters to support it.

Now let us consider the effect of new physics in the decay amplitude.
Since it is possible to obtain the different helicity contributions by
performing an angular analysis \cite{dighe, matias}, from now onward 
we will concentrate
on the longitudinal (i.e., $A_0$ component), which is the dominant one.
In the presence of NP the amplitude can be modified to
\be
A_0 =A_0^{SM}+A_0^{NP}=A_0^{SM}(1+r e^{i \phi_n})\;,
\ee
where $r=|A_0^{NP}/A_0^{SM}|$, and $\phi_n$ is the relative weak
phase between them. For simplicity we set the relative strong phase
between these two amplitudes to zero, which in general is expected to be
small. Thus, in the presence of new physics both in mixing and decay
amplitude the mixing induced CP asymmetry (due to longitudinal
component) is given as
\bea
S_{\phi \phi}&=& 2 ~ \frac{{\rm Im}(e^{-i2(\beta_s+\theta_s) 
} A_0^* \bar A_0)}
{|A_0|^2+|\bar A_0|^2} \nn\\
&=&-\frac{\sin( 2 \theta_s)+2 r \sin (2 \theta_s+\phi_n)+r^2 \sin 
(2 \theta_s+2\phi_n)}{1+r^2+2 r \cos \phi_n}\;.
\eea

To find out the value of $r$ due to unparticle contribution, we now consider
the effective coupling of unparticles to the quarks as 
represented in Eq. (7). Here we consider the effect of
vector like unparticle to the $B_s \to \phi \phi$
decay amplitude. Thus, the transition amplitude due to vector like unparticle
exchange is given as  
\be
A(\overline{B}_s \to \phi \phi)=- e^{-i \phi_{\un}}
\frac{A_{d_{\un}}}{2 \sin d_{\un}\pi}\left
(\frac{m_{B_s}}{\Lambda_{\un}} \right )^{2 d_{\un}-2}
\left (\frac{1}{2}\right )^{d_{\un}-2}
\frac{c_V^{sb} c_V^{ss}}{m_{B_s}^2} 2 X\;,
\ee
where $X$ is the factorized amplitude given in Eq. (\ref{fc}). In the
above equation we have taken the momentum transferred to the unparticle as
$P^2= m_{B_s}^2/2$. Now for numerical evaluation, we use
 a representative value for $c_V^{sb}$, 
i.e., $c_V^{sb}=8 \times 10^{-4}$
from its allowed range, $c_V^{ss}$=0.01 and the same values for 
other parameters  as used in $\Delta M_{B_s}$ case. Thus, we obtain the
the ratio of the NP and SM amplitudes as
\begin{eqnarray*}
r=0.03\;.
\end{eqnarray*} 
It is found that the unparticle contribution to the decay amplitude
is almost negligible.

Now in figure-4, we plot $S_{\phi \phi}$ versus $\phi_n$, the weak phase
in decay amplitude keeping the new mixing phase $\theta_s=20^\circ$
and $\theta_s=0$ (i.e., with no NP contribution to mixing).
From the figure one can see that the unparticle contributions
to the decay amplitude does not have significant effect in $S_{\phi \phi} $.
\begin{figure}[htb]
   \centerline{\epsfysize 2.0 truein \epsfbox{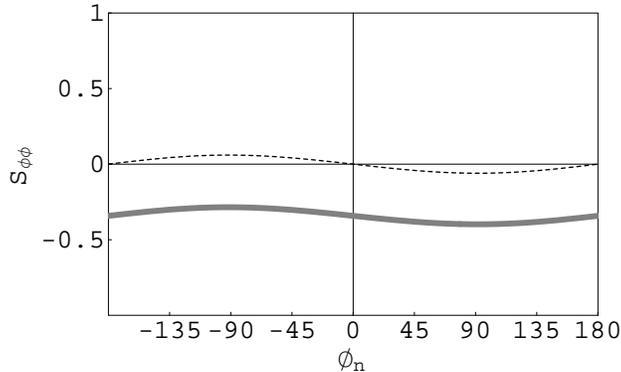}}
 \caption{The variation of $S_{\phi \phi}$ versus
$ \phi_n$ in degree where the thick (dashed) curve is for $\theta_s=20^\circ
(0^\circ)$.}
  \end{figure}

Motivated by the recent proposition of scale invariant unparticle physics
we looked into the effect of the same on the
$B_s - \overline{B}_s$ mixing. In doing so, we included the new physics
contribution, in the form of scalar/vector unparticles, to the SM
contribution and obtained the constraints on the
couplings of unparticle stuff to the SM particles ($c_{S,V}$)
from the data on $ \Delta M_{B_s}$. We found that
due to the effect of  ``unparticles'' large
new mixing phase could indeed be possible. It has also been
observed in Ref. \cite{lenz1} that due to unparticle effects 
it is possible to have large mixing phase in agreement with our
results.

Furthermore, we looked into the possibility of obtaining the mixing induced
CP asymmetry parameters $S_{\psi \phi, ~\phi\phi}$
for the decay modes $B_s \to J/\psi \phi$ and $\phi \phi$, which
can be induced by the new contribution of unparticle stuff. In the SM
the value of $S_{\psi \phi}$ is very small and $S_{\phi \phi}$ is 
identically zero, therefore observation
of non-zero values for these parameters would signal new physics.
Incorporating the NP contribution from the unparticle sector and
using the constraint on $c_{S,V}$ we obtained the values
of $S_{\psi \phi, \phi \phi}$ for the above mentioned 
decay modes, to be nonzero. 
Search for ``unparticle effect'' will be vigorously taken up at the upcoming
experiments and in this context the observation of 
possible large new mixing phase in $B_s - \overline{B}_s$ system and non-zero
values of $S_{\psi \phi,\phi \phi}$
will be very much useful.

\acknowledgments The work of RM was partly supported by Department
of Science and Technology, Government of India, through grant No.
SR/S2/HEP-04/2005. AG would like to thank Council of Scientific and
Industrial Research, Government of India, for financial support.


\begin{thebibliography}{99}

\bibitem{georgi} H. Georgi, Phys. Rev. Lett. {\bf 98}, 221601 (2007);
Phys. Lett. B {\bf 650}, 275 (2007).

\bibitem{kingman} K. Cheung, W. Y. Keung and T. C. Yuan, 
Phys. Rev. Lett. {\bf 99}, 051803 (2007); arXiv:0706.3155 [hep-ph].

\bibitem{ref2} M. Luo and G. Zhu, arXiv:0704.3532 [hep-ph]; C.-H. Chen
and C.-Q. Geng, arXiv:0705.0689 [hep-ph]; arXiv:0706.0850 [hep-ph];
G. J. Ding and M. L. Yan, arXiv:0705.0794 [hep-ph]; 
arXiv:0706.0325 [hep-ph]; Y. Liao,
arXiv:0705.0837 [hep-ph]; C-D.Lu, W. Wang and Y. M. Wang,
arXiv:0705.2909 [hep-ph]; P. J. Fox, A. Rajaraman and Y. Shriman,
arXiv:0705.3092 [hep-ph]; N. Greiner, arXiv:0705.3518 [hep-ph];
D. Choudhury, D. K. Ghosh and Mamta, arXiv:0705.3637 [hep-ph];
S. L. Chen and X. G. He, arXiv:0705.3946 [hep-ph]; 
T. M. Aliev, A.S. Cornell and
N. Gaur, arXiv:0705.1326 [hep-ph]; arXiv:0705.4542 [hep-ph]; 
X.-Q. Li and Z.-T. Wei, 
arXiv:0705.1821 [hep-ph]; P. Mathews and V. Ravindran,
arXiv:0705.4599 [hep-ph]; S. Zhou, arXiv:0706.0302 [hep-ph]; 
Y. Liao and J. H. Liu, arXiv:0706.1284 [hep-ph];
M. Bander, J. L. Feng, A. Rajaraman and Y. Shriman, 
arXiv:0706.2677 [hep-ph]; T. G. Rizzo, arXiv:0706.3025 [hep-ph];
S. L. Chen, X. G. He and H. C. Tsai, arXiv:0707.0187 [hep-ph];
R. Zwicky, arXiv:0707.0677 [hep-ph]; T. Kikuchi and N. Okada,
arXiv:0707.0893 [hep-ph]; C. S. Huang and X. H. Wu, 
arXiv:0707.1268 [hep-ph]; 
D. Choudhury and D. K. Ghosh, arXiv:0707.2074 [hep-ph];
H. Zhang, C. S. Li and Z. Li, arXiv:0707.2132 [hep-ph];
X. Q. Li, Y. Liu and Z. T. Wei, arXiv:0707.2285 [hep-ph]. 

\bibitem{lenz1} A. Lenz, arXiv:0707.1535 [hep-ph].
\bibitem{geng} C.-H. Chen
and C.-Q. Geng, arXiv:0705.0689 [hep-ph].

\bibitem{mass} A.J. Buras, M. Jamin and P.H. Weisz, Nucl. Phys. B
{\bf 347}, 491 (1990).
\bibitem{lim} T.Inami and C.S. Lim, Prog. Theor. Phys. {\bf 65}, 297
( 1981); Erratum- {\it ibid.} {\bf 65}, 1772  (1981).
\bibitem{jl} S. Aoki {\it et al.} [JLQCD Collaboration], Phys. Rev.
Lett. {\bf 91}, 212001 (2003).
\bibitem{hp} A. Gray {\it et al.} [HPQCD Collaboration], Phys. Rev.
Lett. {\bf 95}, 212001 (2005).

\bibitem{ball1} P. Ball and R. Fleischer, Euro. Phys. J. C {\bf 48},
413 (2006).

\bibitem{lenz2} A. Lenz and U. Nierste, hep-ph/0612167.

\bibitem{d0} V. Abazov {\it et al.} [D\O~ Collaboration], 
Phys. Rev. Lett. {\bf 97},
021802 (2006).

\bibitem{cdf} A. Abulencia {\it et al.} 
[CDF Collaboration], Phys. Rev. Lett. {\bf 97},
242003 (2006).
\bibitem{np} M. Blanke, A. J. Buras, D. Guadagnli and C. Tarantino,
J. High Energy Phys. {\bf 10} 003 (2006); Z. Ligeti, M. Papucci and
G. Perez, Phys. Rev. Lett. {\bf 97}, 101801 (2006).
\bibitem{susy} P. Ball, S. Khalil and E. Kou, Phys. Rev. D {\bf 69}, 115011
(2004); S. Khalil, Phys. Rev. D {\bf 74}, 035005 (2006); B. Dutta
and Y. Nimura, Phys. Rev. Lett. {\bf 97}, 241802 (2006); R.
Arnowitt, B. Dutta, B. Hu and S. Oh, Phys. Lett. B {\bf 641}, 305
(2006); X. G. He and G. Valencia, Phys. Rev. D {\bf 74}, 013011
(2006); S. Chang, C. S. Kim and J. Song, J. High Energy Phys. {\bf
0702}, 087 (2007); K. Cheung, C. K. Kang, C. S. Kim and J. Lee,
hep-ph/0702050.

\bibitem{dighe} A. S. Dighe, I. Dunietz, H. J. Lipkin and J. L.
Rosner, Phys. Lett. {\bf B 369}, 144 (1996).


\bibitem{Kra92} G.\ Kramer, W.F.\ Palmer, Phys.\ Rev.\ D {\bf 45} (1992) 193;
G.\ Kramer, W.F.\ Palmer, H.\ Simma, Nucl.\ Phys.\ B
{\bf 428}, 77 (1994).

\bibitem{cheng} Y.H. Chen, H. Y. Cheng and B. Tseng, Phys. Rev. 
D {\bf 59}, 074003 (1999).


\bibitem{ball} P. Ball and R. Zwicky, Phys. Rev. {\bf D 71}, 014029 (2005).

\bibitem{pdg} W. M. Yao {\it et al.}, Particle Data Group, J. Phys.
{\bf G 33}, 1 (2006).


\bibitem{hfag} Heavy Flavor Averaging Group, 
http://www.slac.stanford.edu/xorg/hfag.

\bibitem{matias} S. Descotes-Genon, J. Matias and J. Virto,
arXiv:0705.0477 [hep-ph].
\end{thebibliography}
\end{document}